\documentclass[aps, reprint, superscriptaddress]{revtex4-2}
\usepackage{amsmath,amssymb,graphicx,xcolor,braket,dcolumn,bm}
\usepackage[normalem]{ulem}

\newcommand{\Vb}{V_{\rm b}}
\newcommand{\Ib}{I_{\rm b}}
\newcommand{\Vg}{V_{\rm gate}}
\newcommand{\IV}{I_{\rm b}$($V_{\rm b})}
\newcommand{\VI}{V_{\rm b}$($I_{\rm b})}
\newcommand{\IVa}{I{\text -}V}
\newcommand{\VIa}{V{\text -}I}

\newcommand{\dI}{dI_{\rm b}/dV_{\rm b}}

\newcommand{\ee}{$e$--$e$}
\newcommand{\eh}{$e$--$h$}
\newcommand{\mipt}{Center for Photonics and 2D Materials, Moscow Institute of Physics and Technology, Dolgoprudny, 141700, Russia}

\newcommand{\me}[1]{\textcolor{black}{#1}}
\newcommand{\fig}[1]{\textcolor{magenta}{#1}}
\newcommand{\PP}{6.1} 
\newcommand{\Rv}{8.3} 
\newcommand{\RvMAX}{167} 

\usepackage[utf8]{inputenc}
\usepackage{soul}

\renewcommand{\selectlanguage}[1]{}

\begin{document}

\title{Hysteresis-controlled Van der Waals tunneling infrared detector enabled by selective layer heating}
 
\author{Dmitry A. Mylnikov}
\email{mylnikov.da@yandex.ru}
\affiliation{\mipt}

\author{Mikhail A. Kashchenko}
\affiliation{\mipt}
\affiliation{Programmable Functional Materials Lab, Center for Neurophysics and Neuromorphic Technologies, Moscow 127495}

\author{Ilya V. Safonov}
\affiliation{\mipt}
\affiliation{Programmable Functional Materials Lab, Center for Neurophysics and Neuromorphic Technologies, Moscow 127495}

\author{Kostya S. Novoselov}
\affiliation{Institute for Functional Intelligent
Materials, National University of Singapore, Singapore, 117575, Singapore}

\author{Denis A. Bandurin}
\email{bandurin.d@gmail.com}
\affiliation{Department of Materials Science and Engineering, National University of Singapore, 117575, Singapore}

\author{Alexander I. Chernov}
\affiliation{\mipt}
\affiliation{Russian Quantum Center, Skolkovo Innovation City, Moscow, 121205, Russia}

\author{Dmitry A. Svintsov}
\email{svintcov.da@yandex.ru}
\affiliation{\mipt}

\begin{abstract}

Mid-infrared (mid-IR) photodetectors play a crucial role in various applications, including the development of biomimetic vision systems that emulate neuronal function. However, current mid-IR photodetector technologies are limited by their cost and efficiency. In this work, we demonstrate a new type of photodetector based on a tunnel structure made of two-dimensional materials. The effect manifests when the upper and lower layers of the tunnel structure are heated differently. The photoswitching is threshold-based and represents a ``jump'' in voltage to another branch of the current-voltage characteristic when illuminated at a given current. This mechanism provides enormous photovoltage (0.05$-$1~V) even under weak illumination. Our photodetector has built-in nonlinearity and is therefore an ideal candidate for use in infrared vision neurons. Additionally, using this structure, we demonstrated the possibility of selective heating of layers in a van der Waals stack using mid-IR illumination. This method will allow the study of heat transfer processes between layers of van der Waals structures, opening new avenues in the physics of phonon interactions.

\end{abstract}

\maketitle

\section{Introduction}
Mid-infrared (mid-IR) photodetectors, operating in the 3--15~$\mu$m wavelength range, are invaluable tools across numerous scientific and technological fields \cite{rogalski_infrared_2019}. These devices excel in capturing thermal radiation and revealing molecular vibration lines, enabling diverse applications such as the study of cool celestial objects in astronomy \cite{hinkley_jwst_2022}, non-invasive medical diagnostics through identification of specific volatile organic compounds associated with various health conditions \cite{henderson_laser_2018}, and environmental monitoring by identifying greenhouse gases and pollutants \cite{hodgkinson_optical_2012}. In industry, mid-IR detectors facilitate non-destructive testing and quality control \cite{usamentiaga_infrared_2014}, while in security, they play a crucial role in detecting trace amounts of explosives and chemical agents \cite{bauer_potentials_2008}.

Recently, photodetector devices have found application in the burgeoning field of neuromorphic computing as a part of artificial neurons. Such advancements pave the way for the development of neuromorphic visual sensory and memory systems that can mimic the human brain's ability to process and learn from visual information in the mid-IR spectrum \cite{zhu_hierarchies_2024}. While using separate photodetectors and nonlinear elements as neurons \cite{wu_spike_2020} increases system complexity, devices that combine these functions are more appealing \cite{han_bioinspired_2020}. Despite progress in this field, mid-IR neurons have not yet been demonstrated, with current implementations limited to visible and near-IR wavelengths \cite{luo_artificial_2020,sahu_multifunctional_2023,chen_organic_2023,han_3d_2021}.

A neuron-photodetector must possess nonlinear properties. Superconducting single-photon detectors (SSPDs) exhibit such characteristics and are among the most sensitive IR detectors, capable of distinguishing individual photons. Their sensitivity is primarily due to the disruption of the superconducting state and a sharp voltage jump between $\IVa$ curve branches, which can reach tens of mV. However, they typically operate at ultra-low temperatures below 4 K \cite{walsh_josephson_2021}. Although recent demonstrations of single-photon detection using high-temperature superconductors like BSCCO and LSCO-LCO have been reported \cite{charaev_single-photon_2023}, the operational range of these photodetectors remains limited to the near-IR spectrum \cite{goltsman_middle-infrared_2007,divochiy_single_2018}.

This is where tunnel structures based on two-dimensional materials come into play, having demonstrated potential for IR photodetection \cite{park_large_2024}. Tunnel structures with N-shaped $\IVa$ characteristics closely resemble SSPDs. These structures display negative differential resistance (NDR) in $\IVa$ scans and hysteresis loops with sharp voltage transitions at threshold current values in $\VIa$ scans. The potential to manipulate these transitions through light exposure opens up exciting possibilities for creating switches with significant voltage jumps.

In this paper, we present a threshold detector that harnesses this principle. Our device is a tunnel structure based on a van der Waals stack of graphene/hBN/graphene (Figure \fig{1a}). The band structure of this stack gives rise to multiple resonances dependent on bias and gate voltages, resulting in two distinct NDR regions Figure \fig{1b}). When performing a $\VIa$ scan, we observe hysteresis that shifts under infrared light exposure, producing a significant voltage jump. The threshold switching regime of our device exhibits excitable dynamics and responds to small stimuli. The N-shaped $\IVa$ characteristic of our device, reminiscent of resonant tunneling diodes, enables its operation in a spike generation mode \cite{lourenco_resonant_2022}. This dual functionality as both a detecting and nonlinear element makes it an ideal candidate for photoneuron construction, directly transforming optical signals into electrical ones and thus serving as a true vision neuron \cite{han_bioinspired_2020}.

Intriguingly, this effect manifests when selectively heating either the upper or lower layer of the tunnel structure through targeted illumination. This phenomenon not only demonstrates the device's potential as a neuromorphic component but also presents a novel method for studying vertical heat transfer processes and phonon coupling strengths in van der Waals structures, opening up new avenues for materials science research.

Our theoretical model of tunneling through the hBN barrier fully supports the revealed experimental effects. Using this model, we show that the effects of the current-voltage characteristics can be qualitatively described using the overheating of charge carriers in the place where the laser radiation is focused. It turned out that an essential point in the description of experimental data is the consideration of such effects as the quantum capacitance effect and the transparency dependence of the tunnel barrier on the electron energy. The first one is the main reason for temperature dependence in low temperature regime and the second one describes changing in sign of the thermoresistance at some point.

\begin{figure*}
\includegraphics[width=1\textwidth]{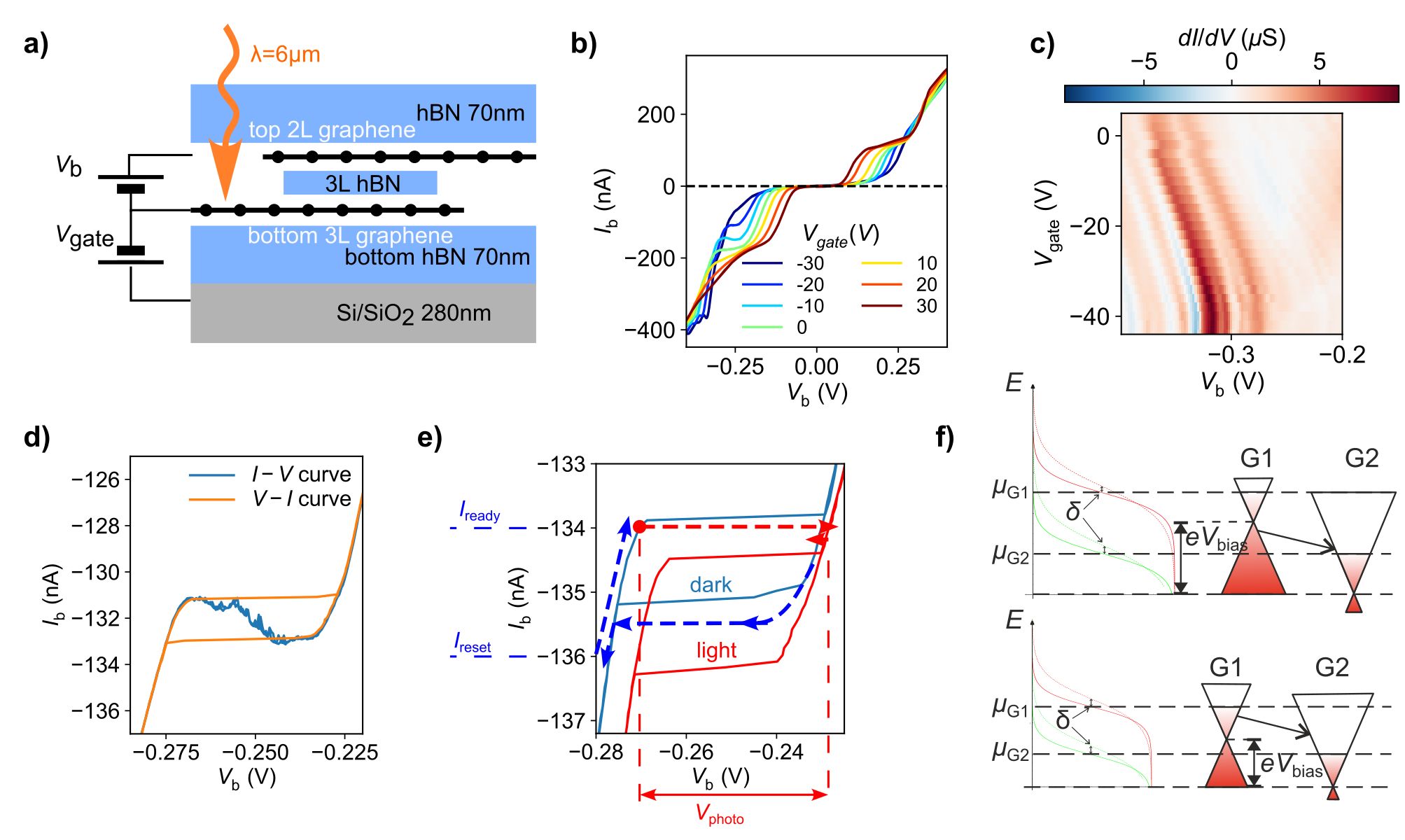}
\caption{Graphene tunneling device and its electrical characterization. a) Stack scheme. b) $\IV$ charachteristics at different gate voltages. c) $\dI$ map on $\Vg$. Two regions of negative differential resistance are visible (blue color). d) $\IV$ charachteristics measured as $I$ on $V$ and $V$ on $I$ at $\Vg=-5$~V. e) $\VI$ without and with illumination and principle of photodetection. Setup schemes are presented in Figures \fig{S2, S3}, Supporting Information, we used scheme \#1 for b,c and \#2 for d,e.  f) Band diagram illustrating effect: Fermi factors of electrons in different graphene layers as functions of energy, illustrations of possible transitions between different layers ($e$--$e$ and $e$--$h$), graphical solution for conservation energy law, equation under the delta function. 
}
\label{fig_electr}
\end{figure*}

\section{Electrical characterization}

First, we observe characteristic ``steps'' in the $\IV$ curve at certain gate voltages. This well-known effect occurs when the Fermi level of one graphene layer aligns with an impurity level in the hBN barrier, creating an additional tunneling channel for carriers via this impurity and leading to a sharp increase in current. Previous studies have extensively investigated the electrical \cite{greenaway_tunnel_2018} and optoelectronic \cite{mylnikov_infrared_2024} properties of impurity-assisted tunneling in the mid infrared range. However, such type of tunneling only leads to monotonic dependence of tunneling current on bias, which does not agree with obtained data.

In our case, we also observe a resonant increase in current at certain bias voltages. As the bias is increased, the current first grows (the resonance), then slightly drops. Two regions of negative differential resistance (NDR) appear at the following gate and bias voltages: $\Vg\approx-10$~V, $\Vb\approx-0.25$~V, and $\Vg\approx-40$~V, $\Vb\approx-0.35$~V (Figure \fig{1c}).

Observing this resonance requires that the top and bottom graphene layers are aligned at a small twist angle. This effect was observed earlier \cite{mishchenko_twist-controlled_2014} and was explained in \cite{brey_coherent_2014}. To describe such behavior we use theory, based on fact of different alignment of graphene sheets relative to each other. This lattice mismatch has crutial role in tunneling processes and divides tunneling mechanisms on two types: transitions between conductance bands of different graphene sheets in small bias regime and transitions between conductance and valence bands after the tunneling resonant. Thus, the occurrence of tunnel resonance and NDR region in $\IVa$ is described by a rapid increase in the current of c-c transitions with bias voltage in comparison with c-v transitions and abrupt decrising of the c-c current after this transitions become forbidden by energy conservation low.

This lattice mismatch has crucial role in tunneling processes and divides tunneling mechanisms into two types: transitions between conductance bands of different graphene sheets in small bias regime and transitions between conductance and valence bands after the tunneling resonance. Thus, the occurrence of tunnel resonance and NDR region in $\IVa$ curve is described by a rapid increase in the current of conduction-to-conduction (c-c) transitions with bias voltage compared to conduction-to-valence (c-v) transitions and an abrupt decreasing of the c-c current after these transitions become forbidden by energy conservation law. In our case, the reason for the small value of hysteresis may be due to the influence of contact resistance \cite{burg_coherent_2017}.

The presence of NDR suggests the existence of hysteresis and abrupt voltage switching. Indeed, when measuring the $\VI$ characteristics, the voltage changes abruptly near these features (Figure \fig{1d}). On the reverse current sweep, the voltage jump occurs at a different current, forming a hysteresis loop. If the $\VI$ characteristic is varied under illumination, a sharp, threshold-like change in voltage can be achieved. While the voltage jump in our sample is around $50$~mV, in samples with smaller twist angles, the typical voltage change can be on the order of volts \cite{mishchenko_twist-controlled_2014, kim_van_2016,burg_coherent_2017}. By biasing the device very close to the hysteresis loop, this voltage jump can be triggered by a very small illumination power. We decided to test this!

\section{Optoelectronic characterization}

\begin{figure*}
\includegraphics[width=1\textwidth]{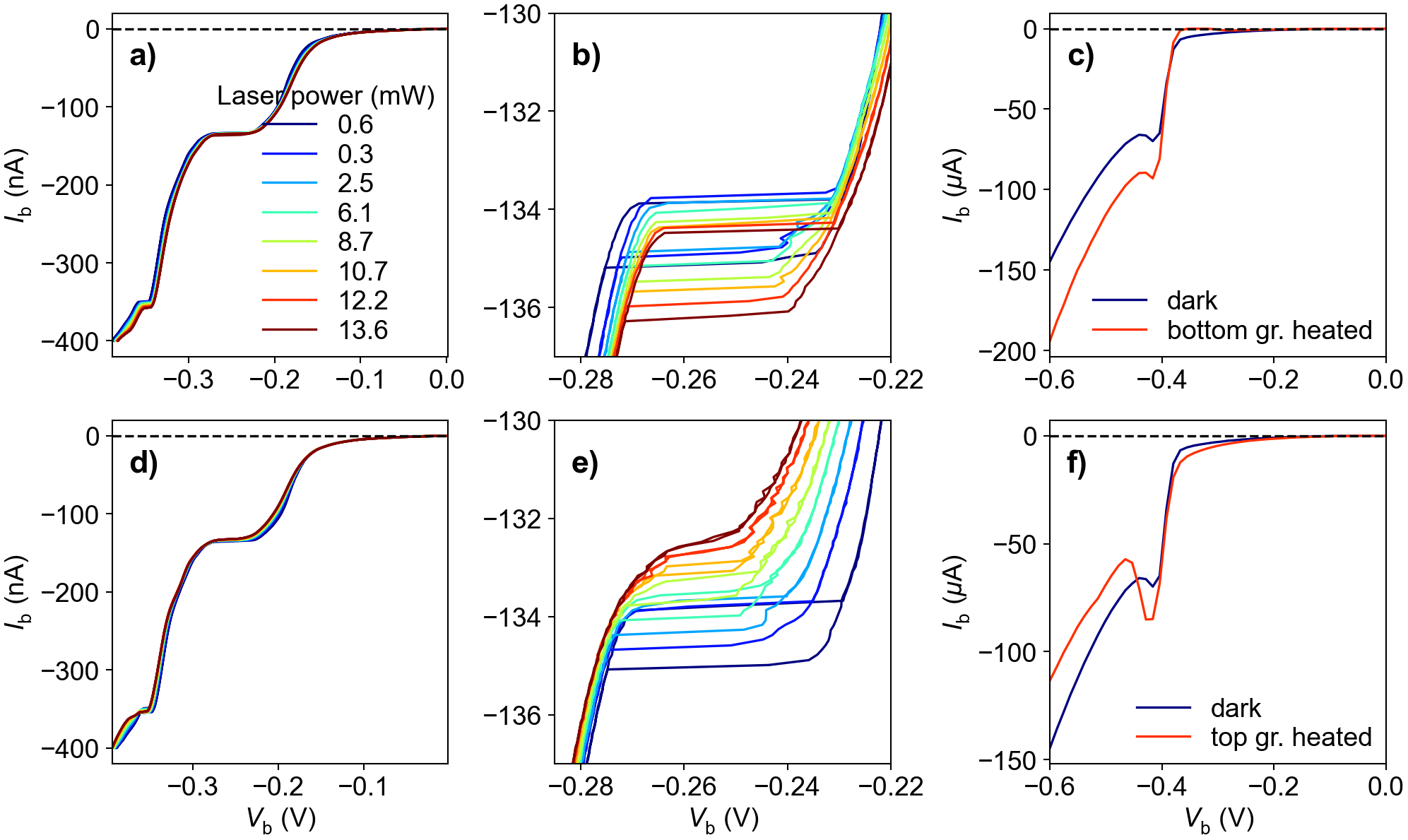}
\caption{$\VI$ characteristics at different illumination power. a) Laser focused at $P$ point, d) at $N$ point. b), e) Hysteresis region on a large scale. c,f) Theoreticaly predicted $\IV$ curves calculated for hBN width of 1~nm, hight of barrier os 1.5~eV \cite{britnell2012electron}, effective mass of 0.5 of electron mass. Blue curve---at 15~K, red---overheated}
\label{fig_IV_laser}
\end{figure*}

\begin{figure*}
\includegraphics[width=1\textwidth]{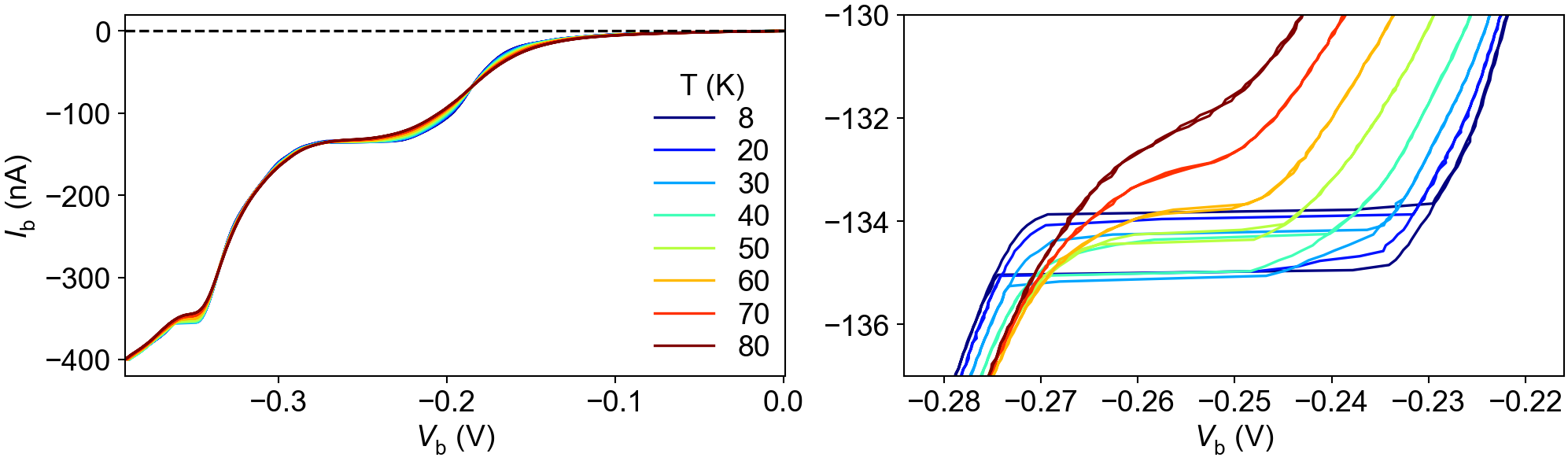}
\caption{$\VI$ characteristics at different sample temperature.}
\label{fig_IV_T}
\end{figure*}

By scanning the sample with a focused laser spot, we identified two spatial regions (labeled as $P$ and $N$) exhibiting markedly different behaviors. Figure \fig{2a,d} shows the $\VI$ characteristics for both regions under increasing laser power illumination. When the $P$-region is illuminated, the temperature dependence is most pronounced near but above the inflection point of the $\IV$ curve at $\Vb=-0.18$~V. In $N$-region, a strong temperature dependence is also observed near the inflection point, but below it and with opposite polarity. Additionally, region $N$ shows strong temperature dependence at $\Vb<-0.3$~V. The temperature-dependent $\VI$ characteristics in regions $P$ and $N$ demonstrate opposite polarities.

The hysteresis loop does indeed shift along the vertical axis under mid-IR illumination, although the temperature dependence of the $\VI$ near the loop is less pronounced. $\VI$ are shown on a larger scale in Figure \fig{2b,e}. In the $P$-region, the current at which hysteresis occurs shifts towards larger negative values under illumination. When the current is fixed at the threshold level, illumination causes a voltage jump in the \textit{positive} direction. That's why we designated this point as the $P$-region. In the $N$-region, the opposite occurs: when illuminated, a voltage jump in the negative direction is observed. This enables threshold switching under light exposure. Although the temperature dependence of $\VI$ characteristics near the hysteresis loop shows minimal shift along the current axis, we observe the maximum voltage jump along the voltage axis.

To demonstrate the threshold photodetector operation, we used a circuit with optical switching and electrical reset of the hysteresis loop. The operating principle for $P$-region illumination is shown in Figure \fig{1e}. The source sets a bias current $I_{\rm ready}$ slightly below the switching threshold. Then the laser is turned on, and under the influence of light the hysteresis loop shifts downwards, triggering a voltage switch of magnitude $V_{\rm photo}$. After the laser is turned off, the current source traverses the hysteresis loop in reverse by first applying $I_{\rm reset}$ and then $I_{\rm ready}$ to return the circuit to its initial state.

\begin{figure*}
\includegraphics[width=0.5\textwidth]{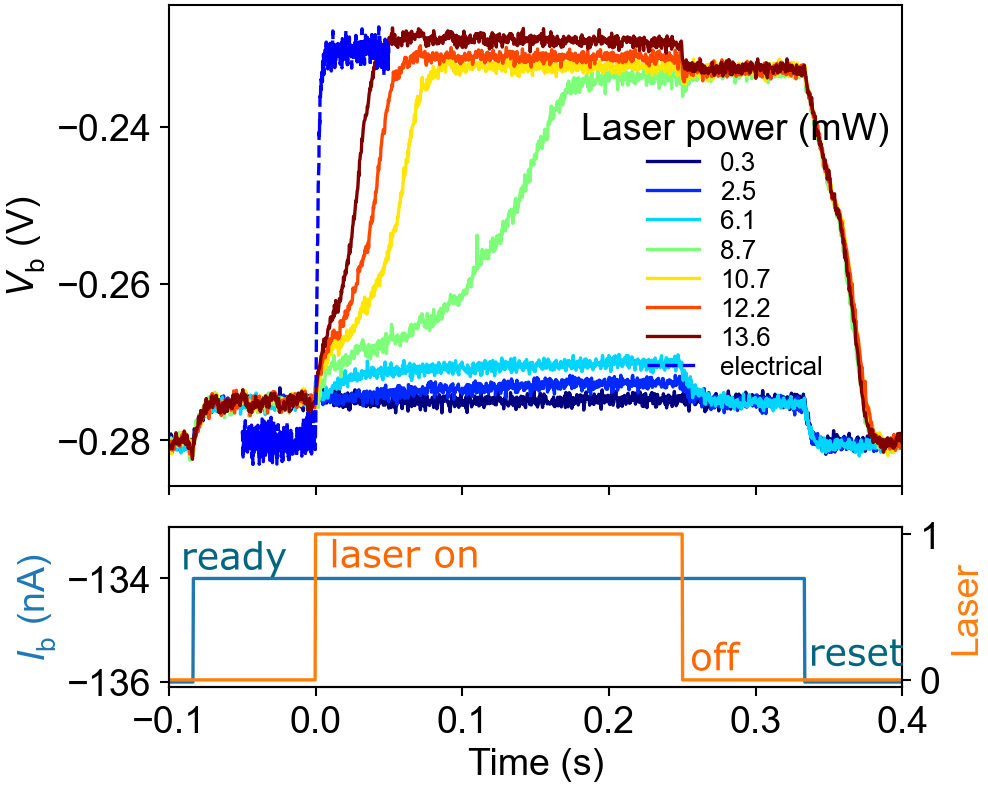}
\caption{Thershold photodetector. a) Waveforms of switching at different illumination power at $P$-point. Ready state: $\Ib=-134$~nA, reset state: $\Ib=-136$~nA.}
\label{fig_hyst}
\end{figure*}

Experimentally measured oscillograms of the threshold detector operation are shown in Figure \fig{4}. In our setup, the cycle repeated with a period $T=0.5$~s. The Keithley sourcemeter set $I_{\rm ready}$, the laser activated after $T/6$ and deactivated after another $3T/6$. After $T/6$, the sourcemeter adjusted the current to reset the hysteresis loop. The oscilloscope and laser were controlled by sync pulses from the sourcemeter.

Switching was measured at various illumination powers and showed threshold behavior. No switching occurred below \PP~mW, while it consistently occurred above this threshold. Higher power resulted in faster switching speeds. Turning on the light causes an abrupt change in voltage on the sample by 50~mV. We could not find illumination power values that would result in a voltage within the middle of the hysteresis loop. Exceeding the power threshold caused an abrupt voltage change across the sample.

To better characterize the effect, we measured the spatial distribution of the photovoltage. We repeated measurements similar to those in Figure \fig{4} at each sample point, with photovoltage measured using a lock-in amplifier at a laser switching period of 0.5~s. Results are shown in Figure \fig{5}. The switching exhibits threshold behavior. At low illumination power, photovoltage is about 5--10~mV, increasing toward the center of the illuminated spot. Above the threshold (\PP~mW), an area appears where photovoltage sharply increases up to $50$~mV and remains constant with further power increases.

The spatial maps visualize the presence of distinct $P$ and $N$ regions discussed earlier. These regions are not only spatially separated but also differ in their hysteresis switching direction under illumination and threshold current values. If we set the current values $I_{\rm ready}$, $I_{\rm reset}$ for switching in the $N$-region, we observe threshold switching in $N$-region, and normal photovoltage in $P$-region. And vice versa - when these currents are set for $P$-region hysteresis, the threshold effect disappears when the laser spot moves to the $N$-region.

The bottom graphs in Figure \fig{5} show switching times measured at each map point. Outside the central spot area, it remains around 5~ms regardless of illumination power. In the central region above threshold power, it increases to 20--30~ms. The fast switching and low photovoltage outside the central spot correspond to ``normal'' photovoltage without hysteresis jump. Similar switching speeds (Figure \fig{S4}, Supporting information) are observed at $\IV$  points outside the hysteresis loop. Hysteresis switching is characterized by larger power-independent photovoltage and slower switching.

Figure \fig{4} includes a dashed line showing electrical switching for comparison. The electrical switching speed is much higher than optical switching and is 1.6~ms. This value aligns with the electrical circuit time constant: $\tau=RC\sim0.4$~ms, where $R=2$~M$\Omega$ is the device resistance, and $C\sim200$~pF represents the combined capacitance of bonds, BNC cables, and instrument input circuits. This also indicates that optical switching speed is not limited by the circuit's $RC$ constant.

\begin{figure*}
\includegraphics[width=1\textwidth]{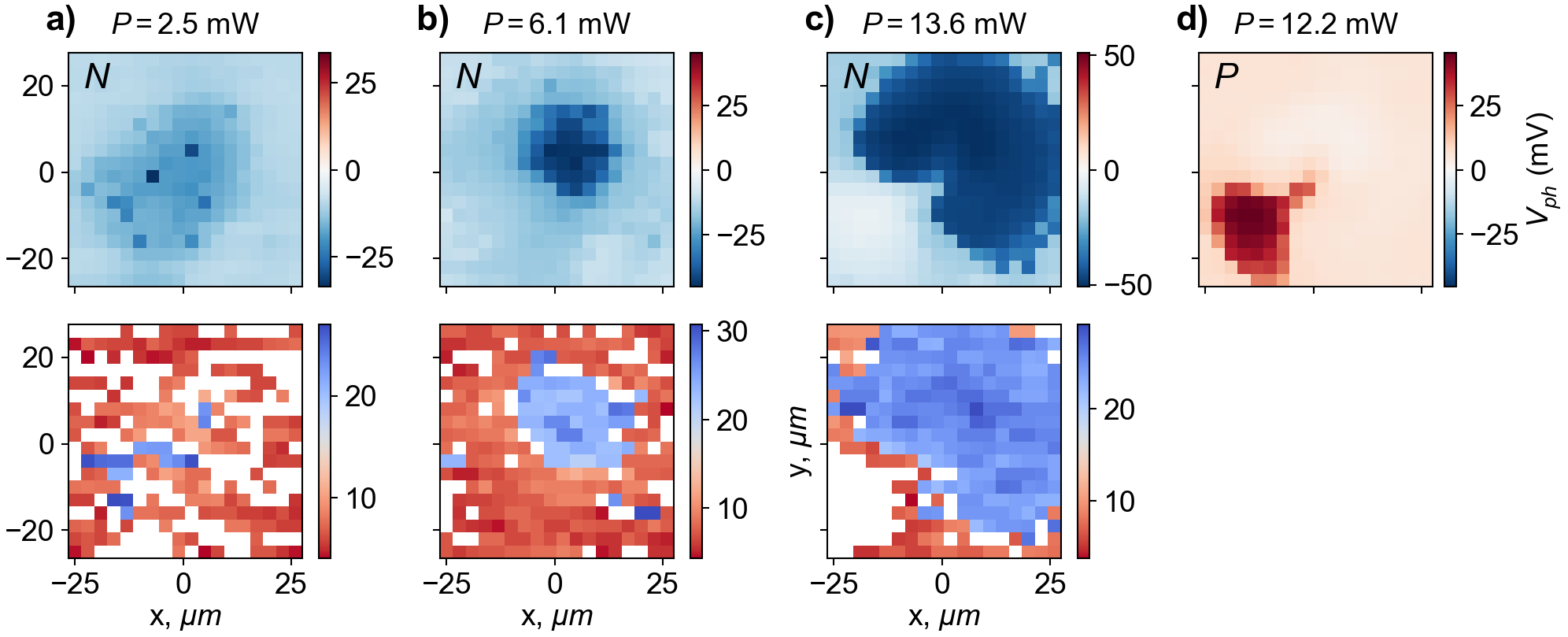}
\caption{Photo switching on hysteresis. a,b,c) Photoswitching when current values $I_{\rm ready}$, $I_{\rm reset}$ are set for switching in the $N$-region at different light powers. Only switching in the $N$-region is observed. d) $I_{\rm ready}$, $I_{\rm reset}$ are set for switching in the $P$-region at high laser power. Top panels show photovoltage, bottom panels show switching time.}
\label{fig_hyst_map}
\end{figure*}

\section{Discussion}

In our structure, we can heat the upper and lower graphene layers separately by radiation, which provides a great opportunity to study the vertical heat transport in van der Waals structures. Let us explain how we discovered this.

Firstly, we recorded the current-voltage $\VI$ characteristics while sweeping the substrate temperature from 8 to 80~K, achieved by changing the cryostat temperature (Figure \fig{3}). The $\VI$ curves exhibit behavior similar to that observed under illumination. Notably, the curves represent the sum of the temperature effects observed previously when heating the sample at points $P$ and $N$. Considering $\VI$ characteristics as a whole, the temperature effect at points $P$ and $N$ is compensated at $\Vb<-0.3$~V, except for the regions of the first and second NDR at $\Vb<-0.35$~V and the region near the inflection point at $\Vb<-0.18$~V. The same is observed in the NDR region. $\VI$ shifts downward near the hysteresis at $\Vb\sim-0.28$~V, corresponding to the behavior at point $P$. $\VI$ shifts upward at $\Vb\sim-0.26$ to $-0.23$~V, similar to the behavior under laser illumination of point $N$.

Secondly, we compared the spatial locations of the $P$ and $N$ regions (Figure \fig{5}) and the positions of the metal contacts (Figure \fig{S1}, Supporting Information). The orientation of the photograph coincides with the orientation of the maps, so the correspondence between the photoswitching regions and the positions of the metal contacts, one of which is connected to the lower graphene and the other to the upper graphene, is clearly visible. At the sharp edges of the contacts, the electromagnetic field is enhanced , leading to increased absorption \cite{nikulin_edge_2021} and heating of the structure.

This clearly indicates that illuminating the $P$ region causes heating of one of the two graphene layers (upper or lower), while illuminating the $N$ region heats the other layer. Heating the whole structure by changing the cryostat temperature leads to the simultaneous heating of both graphene layers, which is seen as the summation of the effects from illuminating the $P$ and $N$ regions. So we have demonstrated a unique way to separately heat the layers in a van der Waals structure, the size of which is smaller than the focused spot size of the radiation.

The response time maps in Figure \fig{5} show a significant difference in photoresponse time outside the hysteresis loop and for switching within the hysteresis. Outside the hysteresis, this is more likely a fast response of hot carriers, in our case probably limited by the $RC$ delay of the circuit. The hysteresis shift, however, is a slower process. We believe that the change in $\IV$ occurs after the hot carriers have transferred their heat to the lattice, mainly by acoustic phonons, since electrons in the temperature range under consideration can excite only this type of phonons. Restricting ourselves to the description of single-electron processes, to excite an optical phonon, there must be a transition with the energy of such a phonon between electron states, but at low temperatures interband transitions are suppressed due to the Pauli principle, and for intraband transitions there will simply be exponentially few electrons with the energy required for the transition. Comparing the changes in $\VI$ upon substrate heating and optical irradiation provides an estimate of the lattice heating of a few tens of Kelvins, which is a typical value for graphene devices.

The effect is very similar to that observed in single-photon detectors. However, the signal magnitude is not record-breaking in our case. Here we have demonstrated the working principle of the threshold detector. Unlike in SSPD (superconducting single-photon detector), we have a large device size, which significantly increases the thermal coupling of graphene and the silicon substrate, reducing the signal. We also illuminated the device with photons of lower energy.

The features observed in the experiment are fully supported by our calculations (see Supporting Information). The current-voltage characteristics predicted by this model are shown in Figure \fig{\ref{fig_IV_laser}c,f}.
 
First, our calculations reproduce two local extrema of $\IV$. They appear when two graphene layers are rotated relative to each other by a small angle. The first one is local maximum and the voltage where it is located correspond to changing types of transitions. Before the critical point current occurs through transitions between conduction band in one layer and also conduction band in another (\ee). After this point dominating transitions become from valence band to conduction band (\eh). The second minimum arises because the electric current from \ee transitions is already rapidly decreasing, and the current of transitions from \eh transitions is increasing, which mathematically leads to the formation of a minimum in the $\IV$-curve.

Second, in full agreement with the experiment, we found that heating one of the graphene layers leads to an increase in the current. Moreover, our calculations predict a change in the sign of the effect in the range of small bias voltages. The position of this point depends on the Fermi energy (change in gate voltage). This behavior results from the dependence of the current on the electron transparency of hBN. Such behavior appears as a result of current dependence on transparency of the hBN for electrons. To understand why this happens, one can refer to the Figure \fig{1f}. In case of \ee{} transitions overheating, when intersection of spectrum cones is above of Fermi energy, leads to increasing number of high energy electrons, which can tunnel the barrier and therefore to increasing of electric current. For these same transitions, but when the intersection is under the Fermi energy, overheating leads to decreasing amount of high energy electrons, which can tunnel through the barrier and therefore to decreasing of current.

Overheating another graphene layer leads to reverse effect (Figure \fig{1f}). This once again confirms our assumption that the illumination of the $P$ and $N$-region corresponds to the preferential heating of the upper or lower graphene separately. The opposite sign of the effect is explained by the Pauli principle: if the number of electrons increases in the vicinity of the intersection of the spectral cones during heating, then there are fewer free places for the transition, therefore the current will decrease, and the opposite situation is for the case where the number of electrons decreases after heating, then the current will increase.

\section{Conclusion}
We have demonstrated a threshold mid-IR photodetector based on a graphene/hBN/graphene tunnel device with NDR and provided theoretical validation for all observed results. Our device, featuring a large misalignment angle and small hysteresis, achieved a sensitivity of \Rv~V/W. For aligned stacks with small rotation angles comprising either two monolayer \cite{mishchenko_twist-controlled_2014} or two bilayer graphene layers \cite{kim_van_2016,burg_coherent_2017}, the sensitivity could reach \RvMAX~V/W due to higher resonance amplitude of up to 1~V. This represents a good performance metric for a mid-IR detector, especially considering that our device was built on a standard silicon substrate without any special measures to enhance light absorption. It's worth noting that we don't normalize the power by the tunnel junction area (6~$\mu$m$^2$), which is much smaller than the illuminated area ($\sim$400~$\mu$m$^2$), since we believe that light absorption and heating occur across the larger area of both top and bottom graphene layers, not just at their overlap in the tunnel junction.

The device's N-shaped $\IVa$ characteristic enables its application as a standalone artificial neuron, eliminating the need for additional nonlinear elements in neuromorphic circuits. This unique combination of mid-IR detection and neuron-like behavior at room temperature represents a significant advance over existing technologies that are limited to visible and near-IR wavelengths or require cryogenic cooling. By bridging the gap between mid-IR photodetection and neuromorphic computing, our work establishes a foundation for developing more efficient and compact bio-inspired visual processing systems. This synergy between mid-IR detection and neuromorphic engineering opens up new possibilities for advanced sensing technologies, from improved medical diagnostics to more sophisticated environmental monitoring systems.

The second significant outcome is our developed technique for selective heating of individual layers in a van der Waals stack. While changing the cryostat's base temperature affects the entire structure uniformly, locally illuminating either the bottom or top graphene region leverages graphene's relatively large dimensions and excellent in-plane thermal conductivity \cite{pop_thermal_2012} to modify the temperature of these layers, including within the small vertical tunnel device area. Our method of scanning the sample with focused mid-IR light provides an excellent approach for detailed investigation of how individual layer temperatures in vertical van der Waals structures affect transport properties---something impossible to achieve through other means. This opens new avenues for studying vertical heat transport in such structures.

\section{Experimental Section}
\textit{Device Fabrication.}
Devices were made using dry transfer technique~\cite{kretinin_electronic_2014}. This involved standard dry-peel technique to obtain graphene and hBN crystals. The flakes were stacked on top of each other (from top hBN to bottom graphite) using a stamp made of PolyBisphenol carbonate (PC) on polydimethylsiloxane (PDMS) and deposited on top of an oxidized (280 nm of SiO$_2$) high-conductivity silicon wafer (KDB-0.001, $\sim$0.001--0.005~$\Omega\cdot$cm). The resulting  thickness of the hBN layers was measured by atomic force microscopy. Then electron-beam lithography and reactive ion etching with SF6  (30 sccm, 125 Watt power) were employed to define contact regions in the obtained hBN/graphene/barrier hBN/graphene/hBN/graphite heterostructure. Metal contacts were made by electron-beam evaporating 3~nm of Ti and 70~nm of Au. The second lithography was done to make a cutout to avoid possible shorting of the top and bottom graphene due to the displacement of the thin hBN layer during transfer. It was followed by reactive ion etching using PMMA as the etching mask. 

\textit{Electrical measurements.}
The sample was held inside a cold finger closed-cycle cryostat (Montana Instruments, s50) at a base temperature of 9~K unless stated otherwise. The differential conductance for Figure \fig{1bc} was measured using alternating current (see Figure \fig{S2}, Supporting Information). A DC bias of $-400$ to $400$~mV from sourcemeter (Keithley Instruments, Keithley 2636B) was mixed with a small AC voltage of 1.5~mV (rms) at 3.335~Hz from a voltage source of lock-in amplifier (Stanford Research, SR860). The same lock-in amplifier was used to measure the AC current through the device. The data were integrated to obtain the $\IV$ characteristic.

Subsequent measurements were conducted in DC mode (see Figure \me{S3}, Supporting Information). The sourcemeter applied DC bias and measured $\IV$ or $\VI$ characteristics. For photomeasurements of hysteresis switching, the sample was illuminated with modulated light. The sourcemeter supplied current through the sample in ``reset'' and ``ready'' states, sending synchronization signals to the laser, lock-in amplifier, and oscilloscope at specified times. In this mode, the AC voltage across the sample was measured using the lock-in amplifier and oscilloscope. When measuring photoswitching maps with the lock-in amplifier, the light signal duty cycle was 50\% at 2~Hz frequency. The measurement circuit employed an SR550 amplifier with gain of 1 to increase the input impedance of the measurement circuit (oscilloscope and lock-in) up to $R_{\mathrm in}=100$ M$\Omega>>R_{\mathrm sample}\sim1$ M$\Omega$).

\textit{IR illumination.}
Linearly polarized light from a quantum cascade laser (Thorlabs Inc.) with a wavelength of 6~$\mu$m was used. Light was focused by ZnSe lens ($D=25$mm) through polypropylene film cryostat window to an almost diffraction limited spot \me{($\sim20\mu$m)}. Motorized XY stage allowed precise aligning of the sample and the laser spot.

\textit{Responsivity calculation}
The amplitude power of modulated laser radiation at threshold ($I=0.35$ A) was 7.6~mW. The power was measured after the focusing lens by two thermal detectors S425C and S302C (Thorlabs Inc.), with the readings averaged. The measurement was made at 2~Hz frequency and multiplied by 2 to obtain power amplitude. Taking into account the measured transmission of the cryostat polypropylene film window at 6~$\mu$m wavelength (0.8), we obtain the power in the focused spot $P_{thr}=\PP$~mW. Thus, the measured device responsivity is $R_V = V_{\rm ph} / P_{thr} = \Rv$ V/W.

\section{Data Availability}
The authors declare that the data supporting the findings are available in the article and its Supporting information. D.A.M. can also provide data upon reasonable request.

\section{Acknowledgments}
The work of D.A.M., M.A.K., I.V.S. and D.A.S. (photocurrent measurements and theoretical modelling) was supported by the Russian Science Foundation, grant \# 21-79-20225. M.A.K. acknowledges the support of an internal grant program at the Center for Neurophysics and Neuromorphic Technologies. The devices were fabricated using the equipment of the Center of Shared Research Facilities (MIPT). D.A.B. acknowledges support from Singapore Ministry of Education Tier 2 grant award T2EP50123-0020. K.S.N. is grateful to the Ministry of Education, Singapore (Research Centre of Excellence award to the Institute for Functional Intelligent Materials, I-FIM, project No. EDUNC-33-18-279-V12) and to the Royal Society (UK, grant number RSRP\textbackslash R\textbackslash190000) for support. 

\section{Author contributions}
D.A.S. and D.A.B. supervised the project; M.A.K., D.A.B. and K.S.N. fabricated the devices; D.A.M. designed the experiment, performed the measurements and analyzed the experimental data with the help from D.A.S., D.A.B. and A.I.C.; I.V.S. and D.A.S. developed the theoretical model; D.A.M. and I.V.S wrote the text with inputs from all authors. All authors contributed to the discussions.

\section{Competing interests}
All authors declare no financial or non-financial competing interests.

\bibliography{zotero, zotero_add}

\end{document}


\title{Supporting Information:\\
Hysteresis-controlled Van der Waals tunneling infrared detector enabled by selective layer heating}
 
\author{Dmitry A. Mylnikov}
\email{mylnikov.da@yandex.ru}
\affiliation{\mipt}

\author{Mikhail A. Kashchenko}
\affiliation{\mipt}
\affiliation{Programmable Functional Materials Lab, Center for Neurophysics and Neuromorphic Technologies, Moscow 127495}

\author{Ilya V. Safonov}
\affiliation{\mipt}
\affiliation{Programmable Functional Materials Lab, Center for Neurophysics and Neuromorphic Technologies, Moscow 127495}

\author{Kostya S. Novoselov}
\affiliation{Institute for Functional Intelligent
Materials, National University of Singapore, Singapore, 117575, Singapore}

\author{Denis A. Bandurin}
\email{bandurin.d@gmail.com}
\affiliation{Department of Materials Science and Engineering, National University of Singapore, 117575, Singapore}

\author{Alexander I. Chernov}
\affiliation{\mipt}
\affiliation{Russian Quantum Center, Skolkovo Innovation City, Moscow, 121205, Russia}

\author{Dmitry A. Svintsov}
\email{svintcov.da@yandex.ru}
\affiliation{\mipt}

\maketitle

\newpage

\section{Supporting Figures}
\begin{figure*}[h]
\includegraphics[width=0.5\textwidth]{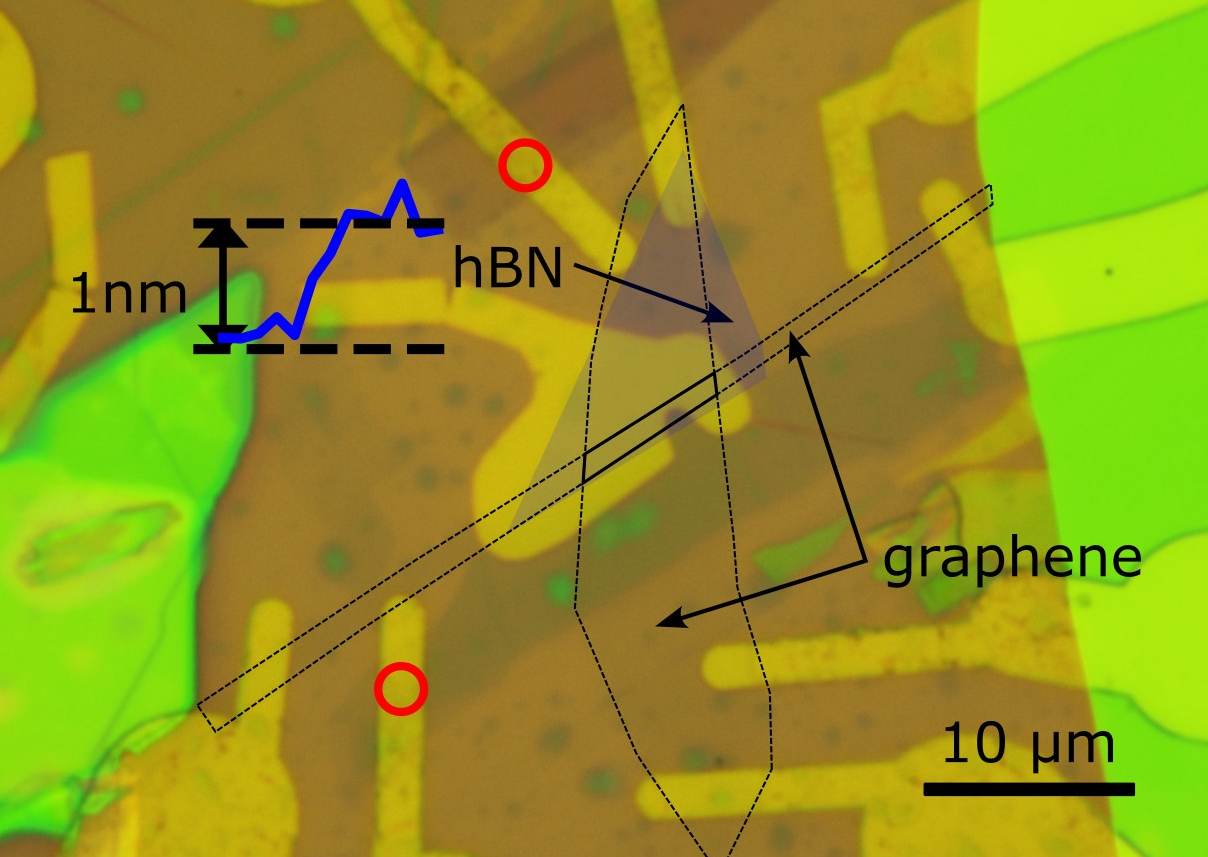}
\caption{Photograph of the sample in the optical microscope with superimposed image of top, bottom graphene and barrier hBN on it. The contacts to the top and bottom graphene used in the measurements are marked in red.}
\label{}
\end{figure*}

\begin{figure*}[h]
\includegraphics[width=0.70\textwidth]{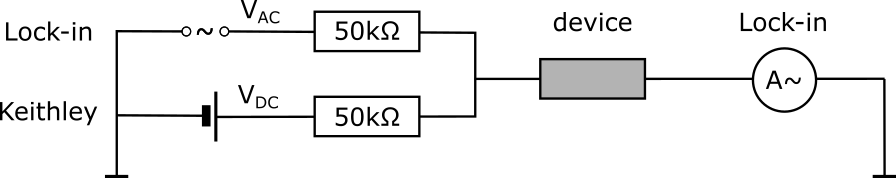}
\caption{Setup scheme \#1. Measuring $\dI$ for Figure 1\me{bc}.} 
\label{fig_Sch1}
\end{figure*}

\begin{figure*}[h!]
\includegraphics[width=0.9\textwidth]{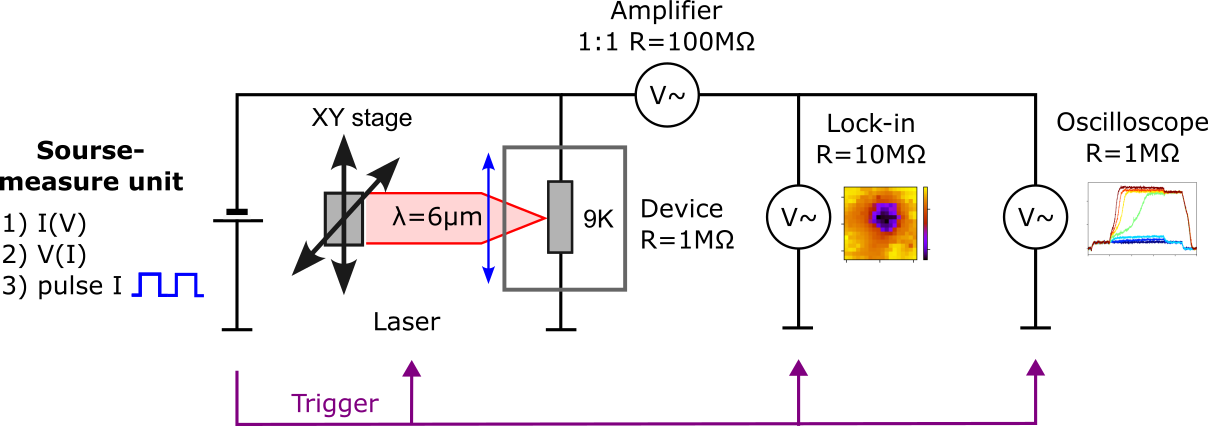}
\caption{Setup scheme \#2. Electrical and optical scheme of the main setup. It was used for $\IV$, $\VI$ measurements, capturing voltage waveforms and photoswitching maps.}
\label{fig_Sch2}
\end{figure*}

\begin{figure*}[h]
\includegraphics[width=0.6\textwidth]{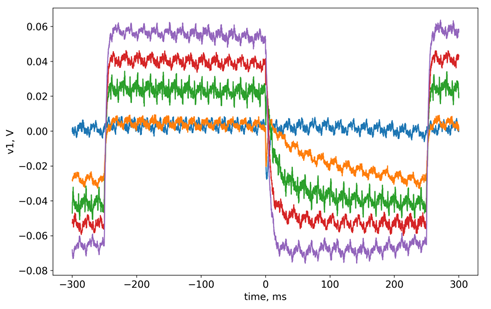}
\caption{Photoswitching is above hysteresis. The rise time is about 5 ms.}
\label{fig_photo_above}
\end{figure*}

\section{Theoretical model}

We develop minimal reproducible theory of tunneling between graphing layers to describe experimental results. As an approximation we consider tunneling between two graphenes. This approach makes theoretical description easier and qualitatively reveals basic properties of such systems. Additionally we take into account that 2-dimensional lattices of graphene sheets rotated relative to each other on the angel $\theta$. This assumption is quite natural, because real devises fabricates by stacking it is complicated to control angel between graphene sheets and therefore one can expect that mismatching between graphene layers is present.

The tunneling current can be estimates as

\begin{equation}
    I=\frac{2\pi e}{\hbar}\int \frac{d k_{x} d k_{y}}{(2\pi)^{2}}D(E)|T_{fi}|^{2}(f_{f}-f_{i})\delta(E_{f}(k_{x},k_{y})-E_{i}(k_{x},k_{y})\pm eV_{d}),
\end{equation}
where integration is over 2D reciprocal space, conservation of momentum is already taken into account, $T_{fi}^{2}=\braket{f|i}\braket{i|f}$ is squared matrix element between final and initial points of tunneling, $f_{f},f{i}$ are Fermi factors of electrons in the initial and final graphen's sheets, $D(E)$ is transparency coefficient, which determine probability to tunnel through the potential barrier. The transparency coefficient in our model estimated by WKB approximation and can be written as
\begin{equation}
    D(E)=exp\left[-\frac{4}{3}\frac{\sqrt{2m_{BN}}}{\hbar}\frac{l}{V_{0}}\Delta_{BN}^{\frac{3}{2}}\left((1-\frac{E}{\Delta_{BN}})^{\frac{3}{2}}-(1-\frac{E}{\Delta_{BN}}-\frac{V_{0}}{\Delta_{BN}}\frac{X_{2}(E)}{l})^{3/2}\right)\right],
\end{equation}
where $X_{2}(E)$ obtains from the equation $E=V(x)$, $V_{0}=Ef_{f}-Ef_{i}+eV_{d}$ is the barrier constant, $V_{d}-$ BIAS voltage between graphene sheets, $\Delta_{BN}, m$ band gap and effective mass for electrons in boron nitride. The spectrum of electrons in rotated graphene defines as
\begin{equation}
    E_{i}=\pm\hbar v_{f}\sqrt{(k_{x}-\frac{\Delta K}{2})^2+k_{y}^2}.
\end{equation}
We choose coordinate system in this way to make specters of electrons in different sheets mode symmetrical.Since the rotation as a whole doesn't change physics of electrons in graphene and only move $K$ point in reciprical space, energy of final state is
\begin{equation}
    E_{f}=\pm\hbar v_{f}\sqrt{(k_{x}+\frac{\Delta K}{2})^2+k_{y}^2},
\end{equation}
where $\Delta K=2\frac{\pi}{a0}\sin(\theta)$ is rotation related correction, that can be obtained by geometrically examining the problem. One can note the symmetry of the equation inside the delta-function. Indeed, the equation
\begin{equation}
    \pm\hbar v_{f}\sqrt{(k_{x}-\frac{\Delta K}{2})^2+k_{y}^2}\pm\hbar v_{f}\sqrt{(k_{x}-\frac{\Delta K}{2})^2+k_{y}^2}\pm eV=0
\end{equation}
defines either a semi-ellipse or one of the hyperbola curves and therefore in the elliptic coordinates equations will be looked more simple and it will be more convenient to estimate the integrals. Thus in elliptical coordinates spectra of electrons in upper and lower graphenes are
\begin{equation}
    |E_{up,low}|=\hbar v_{f} \frac{\Delta K}{2}\left(\cosh{\mu}\pm\cos{\nu}\right),
\end{equation}
where new coordinates define as $k_{x}=\frac{\Delta K}{2}\cosh{\mu}\cos{\nu},k_{x}=\frac{\Delta K}{2}\sinh{\mu}\sin{\nu}$ and matrix elements for electron-electron (ee) and electrons-holes(eh) transitions in this coordinates respectively have the form
\begin{equation}
    <e|e>=\frac{\sinh(\mu)}{\sinh(\mu)+i\sin(\nu)},
\end{equation}
\begin{equation}
    <e|h>=\frac{\sin(\nu)}{\sinh(\mu)+i\sin(\nu)}.
\end{equation}
Final expression for tunneling current for $ee$ and $eh$ transitions with taking account new choose of coordinates
\begin{equation}
    I_{ee}=\frac{2\pi e}{\hbar}\frac{\Delta K^{2}}{16\pi^{2}}\int_{0}^{\infty}d\mu\int_{0}^{2 \pi}d\nu\sinh{\mu}^{2}D(E_{i})(f_{top}-f_{bot})\delta(E_{f}-E_{i}-\Delta),
\end{equation}
\begin{equation}
    I_{eh}=\frac{2\pi e}{\hbar}\frac{\Delta K^{2}}{16\pi^{2}}\int_{0}^{\infty}d\mu\int_{0}^{2 \pi}d\nu\sin{\nu}^{2}D(E_{i})(f_{top}-f_{bot})\delta(E_{f}+E_{i}-\Delta).
\end{equation}

At this stage if we try to estimate this integrals, it will lead us to the divergence in the point, when $\frac{\hbar v_{f}\Delta K}{\Delta}=1$. The reason of such behavior is not taking into account faintness of quasi-particles life-time. Proper calculation requires considering interaction of electrons with phonons and other scattering reasons and it would lead to the different expression for tunneling current. There is an easier way to suppress the divergence by replacing delta function with some narrow bell shape function, which could present itself "non-ideal delta function". Because we already say that we do not estimate finite of quasi-particles life-time properly, we can choose approximation of delta function as gaussian $\delta(x)=A_{norm}e^{-\beta^{2}x^{2}}$.

To estimate tunneling current integrals lets use asymptotic Laplace method for bell shape functions. The reason of applying this approximation is domination of delta function behavior under the other function into the integral. Indeed, the delta function even in "non-ideal" case is very sharp and therefore this property leads to the concentration of under integral function values near the maximum of delta function. The idea of this method is to evaluate area under the integrated function as multiplication of height and characteristic width.

The final form of integrals, which was taken numerically for e-e tunneling 
\begin{equation}
    I_{ee}\approx \frac{e^{2}}{\hbar}\frac{\Delta K^{2}}{8\pi} V_{bias}A_{ee}(V_{bias})\int_{0}^{\infty}d\mu(f_{top}(\mu,\nu_{max})-f_{bot}(\mu,\nu_{max}))\sinh^{2}(\mu),
\end{equation}
where $A(V_{bias})$ is the function which have different dependence in case when energy conservation law has solution and when it doesn't
\begin{equation}
    A_{ee}(V_{bias})=\begin{array}{cc}
         \frac{e^{-\kappa^{2}(1-a)^{2}}}{\sqrt{1-(a-\sqrt{(1-a)^{2}+\frac{1}{\kappa}})^{2}}}&a>1  \\
         \frac{1}{\sqrt{1-(a-\frac{1}{\sqrt{\kappa}})^{2}}}&a<1,
    \end{array}
\end{equation}
where $a=\frac{\Delta}{\hbar v_{f}\Delta K}$ and present itself dimensionless bias voltage. For eh transitions we have similar expressions
\begin{equation}
    I_{eh}\approx \frac{e^{2}}{\hbar}\frac{\Delta K^{2}}{8\pi} V_{bias}A_{eh}(V_{bias})\int_{0}^{\pi}d\nu(f_{top}(\mu_{max},\nu)-f_{bot}(\mu_{max},\nu))\sin^{2}(\nu),
\end{equation}
\begin{equation}
    A_{eh}(V_{bias})=\begin{array}{cc}
         \frac{1}{\sqrt{(a+\frac{1}{\sqrt{\kappa}})^{2}-1}}&a>1  \\
         \frac{e^{-\kappa^{2}(1-a)^{2}}}{\sqrt{(a+\sqrt{(1-a)^{2}+\frac{1}{\kappa}})^{2}-1}}&a<1.
    \end{array}
\end{equation}
The last thing is important to clarify before presenting results is the model for evaluating Fermi energies of considering system. Indeed, the experimental setup has, besides bias, also gate voltage, which additionally controls Fermi energies in graphene sheets. The physical principal by which this system works is called Quantum capacity. Essence of this effect includes taking into account that not all potential difference energy goes to electric field, because some energy went to the changing of Fermi energies in conducting layers. The consideration of this model is similar to the description of an electric capacitor. The solution of the Poisson equation for the electrostatic potential is redundant, since we are interested in the potential difference, because the experiment considers the potential differences between the gate (gate voltage) and between the contacts (bias voltage). Paying off this fact for a simplified system with only one graphene layer and a gate
\begin{equation}
    (\phi_{graphene}-\frac{E_{f}}{e})-\phi_{gate}=V_{gate},
\end{equation}
where $\phi_{graphene},\phi_{gate}$ are potentials on the graphene layer and on the gate respectively. Appearance of term $\frac{Ef}{e}$ in equation above is taking into account effect of quantum capacity: not all potential difference goes to creation electrical field, some of the energy of potential difference goes to the changing of Fermi energy. One more equation to describe this system is taking into account boundary condition for electric field near the conducting graphene layer
\begin{equation}
    E|_{l+0}-E|_{l-0}=4\pi\rho_{graphene},
\end{equation}
where $\rho_{graphene}$ is induced two-dimensional charge density in graphene. Charge density calculates by integrating Fermi distribution, which include dependence on Fermi energy
\begin{equation}
    \rho=-e\left(\int dk^{2}f(E_{e},E_{f})-\int dk^{2}f(E_{h},E_{f})\right),
\end{equation}
where $f(E_{e,h},E_{f})=\frac{1}{\exp{\frac{E_{e,h}-E_{f}}{T}}+1}$ is Fermi factor, $E_{e,h}$ is spectrum for electrons in conductance ($E_{e}$) band or valence ($E_{h}$) band. The last thing is need to do to make system of equations solvable is determine local ground point for potential. In experiment ground point is located on one of the graphene layer, therefore we can do the same of potential in our model, just demand that $\phi_{graphene}=0$. This leads us to $E|_{l+0}=0$ and thus if we take into account that $E_{l-0}=-\phi_{gate}l$ we got solvable system of two non-linear equations with unknown $\phi_{gate}$ and $E_{f}$. Using this approach, it is possible to calculate a system with any number of layers; for our purposes, it is sufficient to apply these considerations only to 2 layers of graphene.